\documentclass[12pt,preprint]{aastex}

\usepackage{epsf}

\def\deg{\ifmmode {^\circ}\else {$^\circ$}\fi}
\def\degree{\ifmmode {^\circ}\else {$^\circ$}\fi}
\def\mum{\ifmmode {\rm \,\mu {\rm m}}\else $\rm \,\mu {\rm m}$\fi}
\def\micron{\ifmmode {\rm \,\mu {\rm m}}\else $\rm \,\mu {\rm m}$\fi}
\def\arcsec{\ifmmode ^{\prime \prime}\else $^{\prime \prime}$\fi}

\def\inch{\ifmmode ^{\prime \prime}\else $^{\prime \prime}$\fi}
\def\Msun{\ifmmode {M_{\odot}}\else $M_{\odot}$\fi}
\def\Mstar{\ifmmode {M_{*}}\else $M_{*}$\fi}
\def\lsun{\ifmmode {\rm L_{\odot}}\else $\rm L_{\odot}$\fi}
\def\mstar{\ifmmode {\rm M_{\star}}\else $\rm M_{\star}$\fi}
\def\lstar{\ifmmode {\rm L_{\star}}\else $\rm L_{\star}$\fi}
\def\md{\ifmmode {M_{\rm d}}\else $M_{\rm d}$\fi}
\def\ld{\ifmmode {\rm L_d}\else $\rm L_d$\fi}
\def\mearth{\ifmmode {\rm M_{\oplus}}\else $\rm M_{\oplus}$\fi}
\def\qdstar{\ifmmode Q_D^\star\else $Q_D^\star$\fi}
\def\kms{\ifmmode {\rm \,km\,s^{-1}}\else $\rm \,km\,~s^{-1}$\fi}
\def\ms{\ifmmode {\rm m~s^{-1}}\else $\rm m~s^{-1}$\fi}
\def\mesc{\ifmmode m_{esc}\else $m_{esc}$\fi}
\def\rmin{\ifmmode r_{min}\else $r_{min}$\fi}
\def\rmax{\ifmmode r_{max}\else $r_{max}$\fi}
\def\mmin{\ifmmode m_{min}\else $m_{min}$\fi}
\def\mmax{\ifmmode m_{max}\else $m_{max}$\fi}
\def\rmind{\ifmmode r_{min,d}\else $r_{min,d}$\fi}
\def\rmaxd{\ifmmode r_{max,d}\else $r_{max,d}$\fi}
\def\mmaxd{\ifmmode m_{max,d}\else $m_{max,d}$\fi}
\def\vrad{\ifmmode v_{rad}\else $v_{rad}$\fi}
\def\qz{\ifmmode q_{0}\else $q_{0}$\fi}
\def\qi{\ifmmode q_{i}\else $q_{i}$\fi}
\def\ql{\ifmmode q_{l}\else $q_{l}$\fi}
\def\qs{\ifmmode q_{s}\else $q_{s}$\fi}
\def\rbrk{\ifmmode r_{brk}\else $r_{brk}$\fi}
\def\rdamp{\ifmmode r_{damp}\else $r_{damp}$\fi}
\def\ain{\ifmmode a_{in}\else $a_{in}$\fi}
\def\aout{\ifmmode a_{out}\else $a_{out}$\fi}
\def\r0{\ifmmode r_{0}\else $r_{0}$\fi}
\def\m0{\ifmmode m_{0}\else $m_{0}$\fi}
\def\M0{\ifmmode M_{0}\else $M_{0}$\fi}
\def\xm{\ifmmode x_{m}\else $x_{m}$\fi}
\def\gyr{\ifmmode {\rm g~yr^{-1}}\else ${\rm g~yr^{-1}}$\fi}
\def\cms{\ifmmode {\rm cm~s^{-1}}\else ${\rm cm~s^{-1}}$\fi}
\def\gcms{\ifmmode {\rm g~cm^{-2}}\else $\rm g~cm^{-2}$\fi}
\def\gcmc{\ifmmode {\rm g~cm^{-3}}\else $\rm g~cm^{-3}$\fi}
\def\pcm{\ifmmode {\rm \,cm^{-1}}\else $\rm \,cm^{-1}$\fi}
\def\psqcm{\ifmmode {\rm \,cm^{-2}}\else $\rm \,cm^{-2}$\fi}
\def\pccm{\ifmmode {\rm \,cm^{-3}}\else $\rm \,cm^{-3}$\fi}
\def\Mdots{\ifmmode {\dot M}_* \else ${\dot M}_*$\fi}
\def\Mdotd{\ifmmode {\dot M}_d \else ${\dot M}_d$\fi}
\def\Msunpery{\ifmmode {M_{\odot}\,{\rm yr}^{-1}} \else $M_{\odot}\,{\rm yr}^{-1}$\fi}
\def\water{\ifmmode {\rm H_2O}\else $\rm H_2O$\fi}
\def\hm{\ifmmode {\rm H_2}\else $\rm H_2$\fi}
\def\ctwohtwo{\ifmmode {\rm C_2H_2}\else $\rm C_2H_2$\fi}
\def\cotwo{\ifmmode {\rm CO_2}\else $\rm CO_2$\fi}
\def\hcop{\ifmmode {\rm HCO^+}\else $\rm HCO^+$\fi}
\def\Lya{\ifmmode {\rm Ly\alpha}\else Ly$\alpha$\fi}

\shorttitle{Disk Sizes and Angular Momentum Transport}
\shortauthors{}

\begin{document}


\title{Protoplanetary Disk Sizes and 
Angular Momentum Transport}


\author{Joan R.\ Najita} 
\affil{National Optical Astronomy Observatory, 950 N. Cherry Avenue, Tucson, AZ 85719} 

\author{Edwin A.\ Bergin} 
\affil{Department of Astronomy, University of Michigan, 1085 S.\ University Avenue, 
Ann Arbor, MI 48109}








\begin{abstract}
In young circumstellar disks, accretion---the inspiral of disk material
onto the central star---is important for both the buildup of stellar
masses and the outcome of planet formation.
Although the existence of accretion is well documented, 
understanding the angular momentum 
transport mechanism that enables disk accretion has proven to be an 
enduring challenge. 
The leading theory to date, 
the magnetorotational instability, 
which redistributes angular momentum within the disk,  
is increasingly questioned, 
and magnetothermal disk winds, which remove angular momentum 
from the disk, have emerged as an alternative 
theoretical solution. 
Here we investigate whether measurements of disk radii 
can provide useful insights into which, if either, of these 
mechanisms drive disk accretion, by searching for evidence of 
viscous spreading in gaseous disks, a potential signature 
of ``in disk'' angular momentum transport. 
We find that the large sizes of most Class II (T Tauri) gas disks 
compared to those of their earlier evolutionary counterparts, 
Class I gas disks, are consistent with expectations for 
viscous spreading in the Class II phase. 
There is, however, a large spread in the sizes of Class II gas disks  
at any age, including 
a population of very small Class II gas disks. Their small sizes may 
result from processes such as photoevaporation, disk winds, or 
truncation by orbiting low mass companions. 

\end{abstract}


\keywords{protoplanetary disks --- accretion, accretion disks ---
stars: variables: T Tauri, Herbig Ae/Be}



\section{Introduction}

Circumstellar disks play a starring role in the formation of 
stars and planets. Stars accrete a significant fraction of 
their mass through disks and planets form from the dust and 
gas in disks. 
Disks surround all stars at birth because the material from 
which stars form, molecular cloud cores, possesses 
more angular momentum than can be contained in the star alone.   
As the disk evolves, the disk material spirals inward 
toward the star, is channeled onto stellar magnetic field lines, 
and eventually crashes onto the stellar surface, producing  
bright ultraviolet (UV) emission. From the luminosity of the UV excess, 
typical (few Myr old) T Tauri stars are inferred to accrete 
at a rate of $\sim 10^{-8}-10^{-7}\Msunpery$ 
(Hartmann et al.\ 1998, 2016), and a $\sim 1\Msun$ star is 
inferred to grow in mass by a few percent to a few tens of percent 
during the T Tauri phase, the initial few Myr of its life. 

Stellar accretion rates are well documented and characterized. 
With measurements now available for hundreds of young stars 
over a range of ages and masses, stellar accretion rates are 
found to decrease with stellar age (Sicilia-Aguilar et al.\ 2010;
Manara et al.\ 2012; Antoniucci et al.\ 2014; Venuti et al.\ 2014), 
increase with stellar mass 
(Muzerolle et al.\ 2003; Calvet 2004; Herczeg \& Hillenbrand 2008; 
Fang et al.\ 2009; Alcala et al.\ 2014; Antoniucci et al.\ 2014; 
Manara et al.\ 2015; Natta et al.\ 2006), and are systematically 
reduced in transition objects, i.e., in disks with large central 
optically thin regions that may 
be forming giant planets (Kim et al.\ 2016; Najita et al.\ 2007, 
2015). 

The inspiral of the accreting disk gas is expected to affect
the outcome of planet formation. 
Giant planets are expected to couple strongly to their gaseous 
disks and migrate inward from their formation distances along 
with the accretion of the disk toward the star.  
The resulting inward Type II migration 
is thought to explain the large number of giant exoplanets that 
are found much closer to their stars than Jupiter is 
in our solar system (e.g., Lin et al.\ 1996).  

Despite the theoretical importance and documented existence of 
accretion, understanding exactly how disk accretion occurs, 
i.e., the mechanism that is responsible for disk angular 
momentum transport,  
has proven to be an enduring challenge.
While the magnetorotational instability (MRI; Balbus \& Hawley 1991) 
had been hailed as the answer to this question for a couple decades, 
recent work finds that non-ideal MHD effects suppress the instability 
in the planet formation region (1--10\,AU), even in the upper disk 
layers, a consequence of the low ionization of T Tauri disks. 
With such ``in disk'' angular momentum transport thus apparently 
suppressed, 
magnetothermal disk winds launched from the disk surface have 
emerged as an alternative angular momentum removal mechanism 
(Bai \& Stone 2013; Kunz \& Lesur 2013; Gressel et al.\ 2015; 
Bai et al.\ 2016; see Turner et al.\ 2014 for a review). 

Angular momentum transport occurs quite differently 
through disk winds and the MRI.  The MRI redistributes  
angular momentum within the disk, so that a small fraction of 
the disk mass acquires most of the angular momentum, which allows 
the rest of the disk to accrete. The disk wind removes angular 
momentum from the disk in order to accomplish the same objective. 
Neither mechanism has a verified observational signature 
thus far, making it difficult to determine which of these, 
if either, drive disk accretion. 

Here we investigate whether measurements of disk radii can provide useful 
insights. If angular momentum transport within the disk is 
important, disks will spread with time as the fraction of the disk 
that takes up the excess angular momentum moves to larger 
radii and the remainder accretes  
(Lynden-Bell \& Pringle 1974; Hartmann et al.\ 1998). 
If disk winds remove the excess angular momentum, disks 
need not grow in size with time. 

Previous commentary on this topic has largely focused on 
the possible change in size with age of the 
{\it dust} component of Class II disks. 
In star-forming regions separated by a few Myr in age, 
the dust component of disks, as measured from submillimeter 
continuum emission, is found to be slightly larger for 
older disks in Lupus than those associated with the younger 
Taurus and Ophiuchus populations (Tazzari et al.\ 2017).  
While one might hope to detect more obvious evolution 
in disk size by comparing disk size measurements at 1--3 Myr to 
those of even older populations ($\sim 10$ Myr), 
disk photoevaporation induced 
by stellar FUV irradiation has been argued to significantly reduce 
the size of a gaseous disk on few Myr timescales 
(e.g., Gorti et al.\ 2015). Photoevaporation will also strip 
away small grains that are coupled to the gas, potentially 
making the effect of 
viscous spreading difficult to detect at late times. 
Moreover, the likelihood that the large grains responsible for 
disk submillimeter emission migrate inward early in 
the evolution of disks (Takeuchi \& Lin 2002, 2005; 
Birnstiel \& Andrews 2014) suggests that submillimeter 
continuum measurements will underestimate the radii 
of Class II gas disks (e.g., Ansdell et al.\ 2018).  

To sidestep these difficulties, here we compare the evolution 
in the size of {\it gaseous} disks, focusing on the evolution 
at earlier times, between the Class I and Class II phases. 
Section 2 describes the observational data that we use to address 
this issue. Sections 3 and 4 describe our result and its implications.

\section{Methods and Data}

To place an observational constraint on the 
mechanism that transports angular momentum in the T Tauri phase, 
we focus on the disk radii of Class I and Class II sources. 
Class I sources are young stellar objects 
that are still embedded in their molecular envelope. 
Stars accrete much of their mass during the Class I phase. 
Fed by infall from the molecular envelope, 
their surrounding disks are expected to be massive, with   
disk accretion likely to be driven by 
gravitational instability and possibly episodic in nature 
(Hartmann et al.\ 2016; Zhu et al.\ 2010). 
Class II sources are evolutionarily older: their infall 
having ceased and their molecular envelopes dissipated, 
their disk mass declines with time as 
the star accretes through the disk at a more leisurely pace. 

If the sizes of Class I disks establish the ``initial'' disk 
sizes at the beginning of the Class II phase,  
we can infer whether disks spread or not in the Class II phase 
as a consequence of accretion 
by comparing the sizes of Class I and Class II disks. 
The sizes of Class I disks will be affected by several factors: 
the angular momentum of the cloud core from which it formed, 
any magnetic braking that occurs in the collapse and infall 
process, as well as any spreading that occurs through 
disk angular momentum transport in the Class I phase. 
The sizes of Class II disks can also increase through viscous spreading 
as well as decrease through the action of FUV-driven photoevaporation
and truncation by orbiting companions. 

In measuring the size of a rotationally-supported disk in a Class I source, 
it is important to distinguish the disk emission from that of the 
infalling envelope.  
In contrast to a rotationally-supported disk, which will show 
a velocity trend of $v_\phi(r) \propto r^{-1/2}$ for Keplerian rotation, 
infalling gas at larger 
radii that conserves angular momentum will follow 
$v_\phi(r) \propto r^{-1}.$
Harsono et al.\ (2013) argued that Class I disk sizes could be determined 
by searching for the radius where the velocity field transitions 
from $v_\phi \propto r^{-1}$ to $v_\phi \propto r^{-1/2}$. 
High angular resolution is needed to resolve the velocity field into these 
components.

Table 1 lists the properties of Class I sources with reported 
disk sizes that are derived in the above way from spatially resolved 
emission.
The molecular tracers used are primarily CO and its isotopes, 
as well as HCO$^+$ and CS.
Because of the small number of Class I sources studied to date, we have 
also included Class 0 sources that have accretion rates and 
stellar masses similar to those of the Class I sources 
in Table 1. 
The stellar masses listed for these sources (referred to 
hereafter generically as ``Class I sources'') 
are derived dynamically from the spatially resolved velocity structure 
of the disk. 

To select Class I sources that are the plausible precursors of Class II 
T Tauri stars,
we selected sources with stellar masses in the same range as 
the Class II sources (0.3--1.3$\Msun$; see below). 
While the stellar masses of Class I sources will grow through accretion 
before reaching the Class II stage, including Class I sources 
in the same mass range as the Class II sources allows for the 
possibility that these sources 
are at the end of the Class I phase. 
This is a conservative choice in that more massive protostars 
tend to have larger disks. 
Note that several sources in 
Table 1 may evolve into Class II sources with stellar masses 
above 1.3\Msun\ if they continue to accrete at their current 
rates for $\gtrsim 0.2$\,Myr (e.g., L1551 IRS5; HH 212). 

Because the sample size is small, 
IRS 63 is included for completeness, despite the fact that its 
near face-on geometry makes it difficult to distinguish between 
its disk and envelope (Brinch \& Jorgensen 2013).
Two sources that are well-known protostellar binaries 
(L1551 NE and L1551 IRS5) are also included for completeness, 
although their circumbinary disks may be preferentially larger 
than than disks associated with single stars, e.g., if the system 
formed from high angular momentum material. 

As shown in Table 1, the rotationally-supported disks of 
Class I sources with central stellar masses $< 1.0 \Msun$ 
have disk radii $R_d$ from 50\,AU to 300\,AU 
with a typical value of $\sim 100$\,AU.

\begin{deluxetable}{lllllll}
\tabletypesize{\footnotesize}
\tablecaption{\label{t:std} Properties of Class 0 and I Sources}
\tablehead{
Name	    & $\Mstar$  & $\Mdots$ & $R_{\rm gas}$ & $\md$    & References \\
    	    & ($\Msun$) & ($\Msunpery$) & (AU)	& ($\Msun$)    & }
\startdata
TMC-1A      & 0.64      & 4(-7) & 100   & 0.04  & Aso et al.\ (2015) \\
L1551-IRS5  & 0.5       & 4(-6) & 64    &       & Chou et al.\ (2014) \\
HH212	    & 0.2	& 8(-6) & 120   &       & Lee et al.\ (2014) \\
L1527       & 0.3       & 6(-7) & 54    &       & Ohashi et al.\ (2014) \\
L1551 NE    & 0.8       & 5(-7) & 300   &       & Takakuwa et al.\ (2012) \\
IRS 63	    & 0.8       & 1(-7)	& 170	& 0.1   & Brinch \& Jorgensen (2013) \\
TMC1	    & 0.54	& 2(-7)	& 100	& 0.037	& Harsono et al.\ (2014) \\
TMR1	    & 0.7	&	& 50	& 0.012	& Harsono et al.\ (2014) \\
L1536	    & 0.4	&	& 80	& 0.015	& Harsono et al.\ (2014) \\
Elias 2-27  & 0.55	&	& 300	& $\sim$0.1	& Perez et al.\ (2016), Tomida et al.\ (2016) \\
Lupus 3 MMS & 0.3       & 1(-7) & 130   &       & Yen et al.\ (2017) \\
L1455 IRS1  & 0.28      & 1(-6) & 200   &       & Harsono et al.\ (2014) \\
VLA 1623    & 0.2       & 6(-7) & 150   &       & Murillo et al.\ (2013) \\
\enddata
\vspace{-0.5cm}
\end{deluxetable}

To measure the size of Class II disks, 
we also use tracers of the gaseous component of disks rather 
than the dust.  
Disk solids are expected to experience significant inward 
radial drift relative to the gas (e.g., Takeuchi \& Lin 2002, 2005; 
Brauer et al.\ 2008; Birnstiel \& Andrews 2014). As a result, 
the size of the dust continuum emission underestimates 
the true disk size 
(e.g., Andrews et al.\ 2012; Huang et al.\ 2018; 
Liu et al.\ 2017; Ansdell et al.\ 2018) 
and gaseous tracers are preferred.

Good tracers of the radial extent of the gaseous disk 
would separate the disk emission from that of any surrounding 
molecular cloud. 
While estimates of disk radii made from CO emission work well 
in the absence of a surrounding molecular cloud, 
CN $N$=2--1 can be a better choice when a cloud is present. 
In their study of disks in the Taurus star forming region, 
Guilloteau et al.\ (2013; hereafter G13) showed that 
the CN $N$=2--1 emission from disks 
experiences little cloud contamination,  
is strong enough to be commonly detected, 
and is well-behaved. 

Here we focus on gas disk sizes measured for Taurus because 
it is a relatively young star forming region (1--2\,Myr), 
is well studied, and has limited contamination from the 
molecular cloud. 
We collated from the literature gas disk sizes for 
Class II sources 
with consistently determined mass and age estimates 
(Andrews et al.\ 2013),
excluding sources with a spatially resolved 
stellar companion within $2\arcsec.$ 
Stellar companions with a separation comparable to the 
disk size can truncate the outer radius of the disk 
(e.g., Artymowicz \& Lubow 1994). As a result, 
sources such as RW Aur ($1.4\arcsec$), UY Aur ($0.88\arcsec$), 
T Tau ($0.7\arcsec$) are excluded.
Because companions on scales much smaller than the disk
($\ll 1\arcsec$; e.g., GG Tau, DQ Tau) would not 
truncate the outer disk radius, systems with such 
companions are included.

As part of a study that used 
the Keplerian rotation signature of the CN $N$=2--1 
emission from disks to derive stellar masses,  
Guilloteau et al.\ (2014; hereafter G14) 
reported gas disk outer radii determined from 
a power law fit to the emission that extends to 
an outer radius $R_{\rm out}$, 
as derived from the IRAM interferometric data.
The reported gas disk outer radii range from 225--750 AU. 
Guilloteau et al.\ (2016; hereafter G16) collated these 
and other gas disk sizes reported in the literature 
for Taurus sources, as measured interferometrically  
using the tracers CO, $^{13}$CO, CN, and HCO$^+$.  
Most of the interferometric radii in Table 2 
($R_g$, column 7) are from this compilation, 
supplemented by measurements from Simon et al.\ (2017). 
The errors on gas disk size reported in the literature 
range from 1\% to 33\% with a median error of 7\% for the 
sources included in our study. 

When spatially resolved data are unavailable, 
line fluxes $F$ 
can be used instead to estimate the radial extent 
of the disk gas $R_{\rm out}$ 
when the system inclination is known.  
As described by G13, $R_{\rm out}$ can be derived using 
the relation
$F = B_\nu(T)(\rho \Delta V)\pi R_{\rm out}^2/D^2\cos(i)$, 
where $T$ is the average disk teperature, $\Delta V$ is the 
local line width, $D$ is the distance, and $\rho$ is a dimensionless 
parameter that depends on the line opacity. 
Disk sizes have been measured with this approach 
using CN $N$=2--1 emission (G13) 
and 
\hcop\ $J$=3--2 emission (G16) 
assuming $\Delta V = 0.2\kms$ and  $T$=15\,K.  
The line flux-based CN radii in Table 2 
($R_{\rm CN}$, column 6) are from G13.
As discussed by G16, 
the derived outer radii are very similar to (within 20\% of) 
sizes obtained from resolved images 
made with interferometers, where available (see also Table 2). 

As interferometric disk radii are available for most of 
the sources in Table 2, in our study we adopt these 
in lieu of the line flux-based sizes when available. 
The line flux-based CN radii are adopted for 6 sources, 
4 of which are upper limits 
(DO Tau, FT Tau; upper limits for BP Tau, CIDA-11, DQ Tau, DR Tau).

For comparison with the radial extent of the gaseous 
emission, Table 2 also collates from the literature
dust disk sizes, where available.
Most of the values are from the study of 
Tripathi et al.\ (2017), who report dust disk sizes measured 
interferometrically from 
data obtained with the Submillimeter Array (SMA). 
Andrews \& Williams (2007) report outer radii for the dust 
continuum emission for Taurus T Tauri stars. 
Other continuum sizes have been reported 
by Pietu et al.\ (2014) and Harris et al.\ (2012). 

The outer dust disk radii reported by Pietu et al.\ (2014), 
based on data from the IRAM Plateau de Bure interferometer, 
come from fitting the visibilities to a parametric model 
that assumes a power law surface density distribution 
$\Sigma(r) \propto r^{-p}$ for disk radii $r$ out to 
an outer radius $R_{\rm out}$ and a power law index $p=1$.  
As they note, the derived size is relatively insensitive 
to the choice of $p$.  Disk radii would be 
$\sim 15$\% smaller for $p=0$ and 
$\sim 5$\% larger for $p=2.$
The outer disk radii reported by Harris et al.\ (2012), 
based on data taken with the Submillimeter Array (SMA),  
were similarly obtained by fitting the disk emission to 
a simple parametric model with surface brightness 
$I_\nu \propto r^{-1.5}$ out to an outer radius $R_d$. 
They estimate that altering the power law index by 
$\pm 30$\% would change the disk size by 
$\sim 20$\% to 40\%, with steeper (shallower) gradients 
corresponding to larger (smaller) disk sizes. 

Tripathi et al.\ (2017) model the SMA visibility data 
using a more complex 5-parameter ``Nuker'' profile for 
the intensity, $I_\nu(r) \propto (r/t_t)^{-\gamma}
[1 + (r/r_t)^\alpha]^{(\gamma-\beta)/\alpha},$
that is capable of fitting the wide range of morphologies 
of disks, from continuous disks to the ring-like 
emission of transition objects. From the fits, one 
can retrieve the emission size of the disk $R_{\rm eff}$ 
that encompasses a fixed fraction of the total flux. 
Although Tripathi et al.\ (2017) tabulate disk sizes that 
enclose 68\% of the total flux, here we use the corresponding 
values that enclose 90\% of the total flux, which more 
closely captures the radial extent of the disk continuum 
emission. 

The errors on dust disk size reported by Tripathi et al.\ 
are 1\% to 13\% with a median of 5\% for the sources included 
here. Pietu et al.\ (2017) reported larger errors (5\% to 
27\%) on dust disk size, with a median of 11\% for the 
sources included in our study.

\begin{figure}[tbh!]
\figurenum{1}
\includegraphics[height=3in]{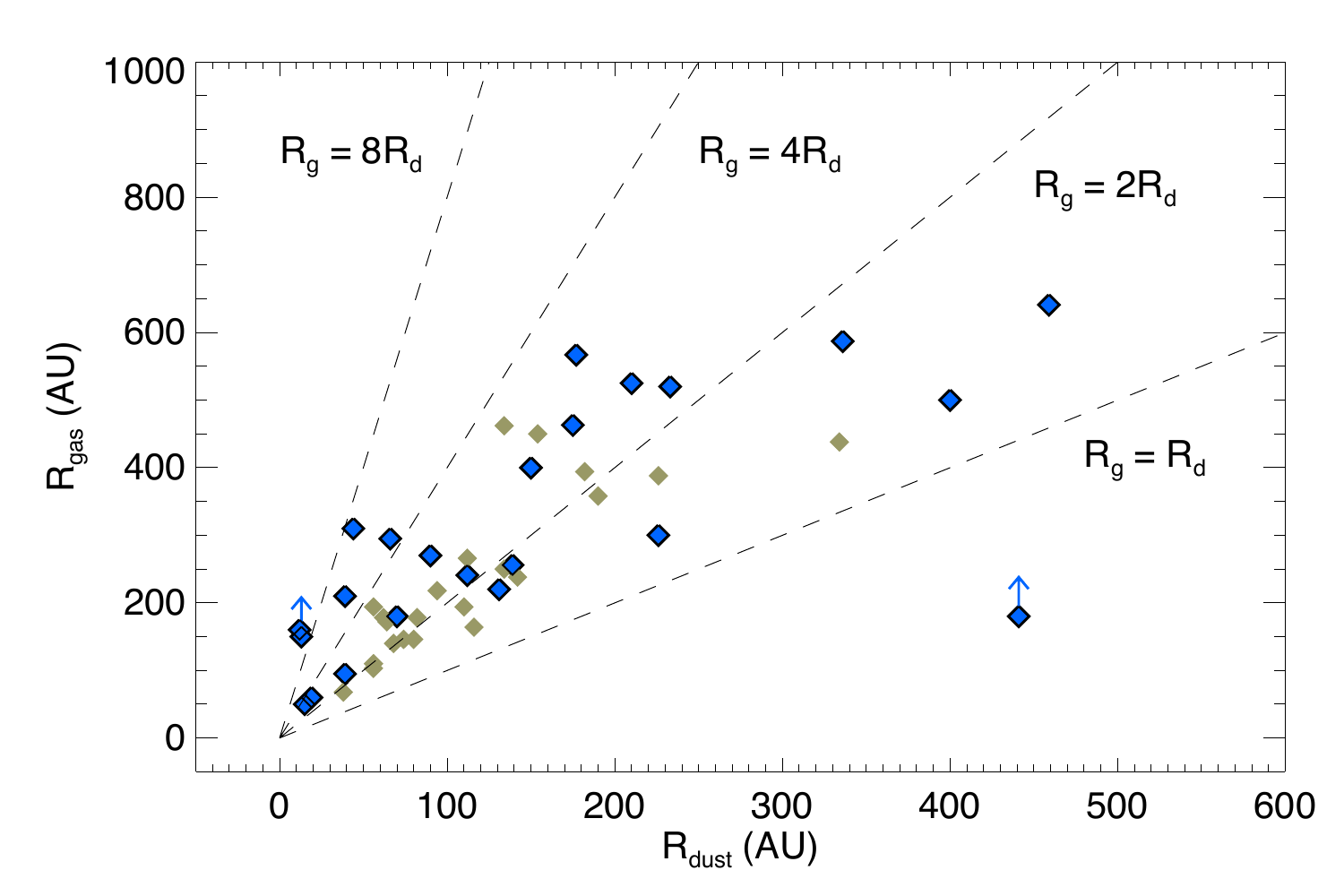}
\caption[]
{\small Sizes of Class II disks in Taurus (large blue 
diamonds) as measured 
interferometrically from dust continuum emission 
and gaseous tracers, 
for all sources where both measurements are available.
Sources have gas disks typically 1.5 to 8 times larger than 
their dust disks. Many gas disks are larger than $200$\,AU.
Similar results reported by Ansdell et al.\ (2018) are also 
shown (small gray diamonds).
}
\end{figure}

\begin{figure}[tbh!]
\figurenum{2}
\includegraphics[height=4in]{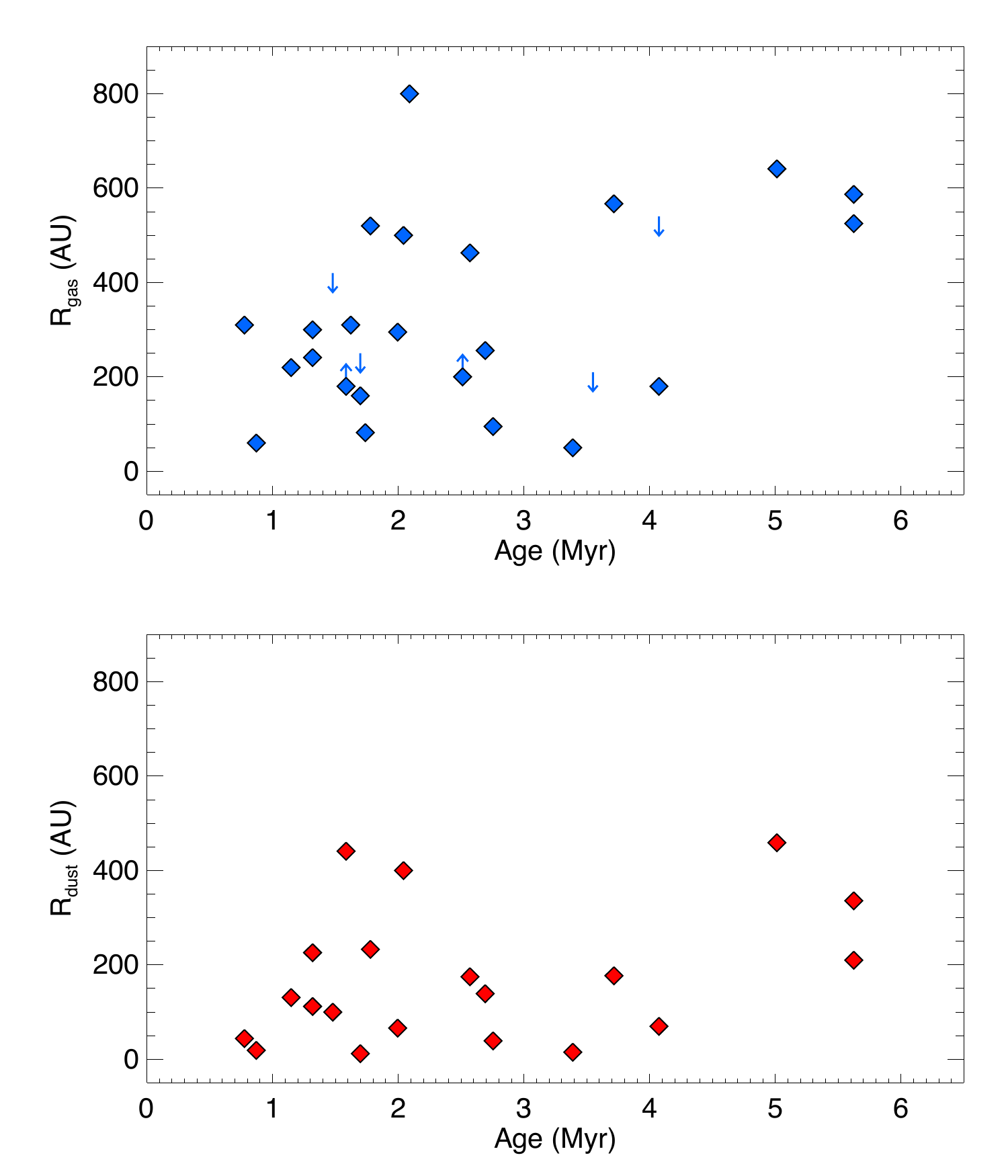}
\caption[]
{\small Sizes of gas dust disks of 
Taurus classical T Tauri stars as a function of stellar age. 
Gas disk sizes (top) are 
measured from interferometrically for the vast majority 
of sources (blue diamonds) 
or have an upper limits from CN line fluxes (blue downward 
arrows). 
Dust disk sizes (bottom) are measured interferometrically from 
submillimeter continuum emission. 

}
\end{figure}

\section{Results} 

Figure 1 compares the millimeter continuum sizes of 
dust disks and gaseous disks, both measured 
interferometrically, for sources where both measurements 
are available. 
As found in other studies of individual sources 
(e.g., Andrews et al.\ 2012; Huang et al.\ 2018; 
Liu et al.\ 2017; see also 
Isella et al. 2007; Panic et al.\ 2009; 
de Gregorio-Monsalvo et al.\ 2013; Cleeves et al.\ 2016) 
and star formation regions 
(Ansdell et al.\ 2018), 
the gaseous 
component of disks extends to larger radii than the dust 
component. In this sample, sources have gas disks typically 
1.5 to 8 times larger than their dust disks, with most 
gas disks larger than 200\,AU. 
The larger sizes of Class II gas disks compared to 
Class II dust disks are likely the result of 
inward radial drift of dust through aerodynamic drag.  

Similar to the situation found here for the Taurus disks, 
Ansdell et al.\ (2018) 
reported that the gas disks of 22 sources in the Lupus 
star forming region 
are uniformly larger than their dust disks by a factor of 
1.5 to 3. 
As they describe, the difference could be due to either  
the inward drift of disk solids relative to the gas 
and/or optically thick gas emission. 
In the latter scenario, 
the CO emission is optically thick,  
making it easier to detect at larger radii than 
optically thin dust continuum.  
Ansdell et al.\ (2018) found that while optical depth effects 
could account for the lower $R_g/R_d$ values they observed, 
radial drift was needed to explain the higher observed
values.  
In either case, the true radial extent of the disk is 
best probed with a gaseous tracer.

Figure 2 plots the gas and dust disk sizes as a function 
of stellar age (Andrews et al.\ 2013, using the Siess et al.\ 2000 tracks) 
for all sources from Table 2 in a stellar mass range 
(0.3--1.3\,$\Msun$) appropriate for the 
evolutionary descendants of the Class I sources 
in Table 1. That is, we exclude both very low mass stars  
(CIDA-1, CIDA-8, FN Tau, FP Tau)
as well as intermediate mass T Tauri stars and Herbig 
stars (AB Aur, CW Tau, MWC758, RY Tau, SU Aur, and T Tau).
The gas disk sizes shown are measured interferometrically 
for the vast majority of sources 
(blue diamonds) or have an upper limit from their CN line flux 
(downward blue arrows; G13). 
There is a large range in disk size at any age and 
no strong trend of size with age, although 
sources with gas disks larger than 400 AU are older than 1.5 Myr. 

The gaseous disk sizes in Figure 2 include measurements made 
with CN, which may underestimate the gas disk size. 
Because CN $N$=2--1 can be subthermally populated, it  
may not trace the full extent of the gas disk. 
Guilloteau et al.\ (2016) find that \hcop\ $J$=3--2, 
which is better thermalized than CN, is optically thick  
and a good measure of the extent of the gas disk.
When both diagnostics are measured, the CN size is often smaller. 

\begin{figure}[h!]
\figurenum{3}
\includegraphics[height=3in]{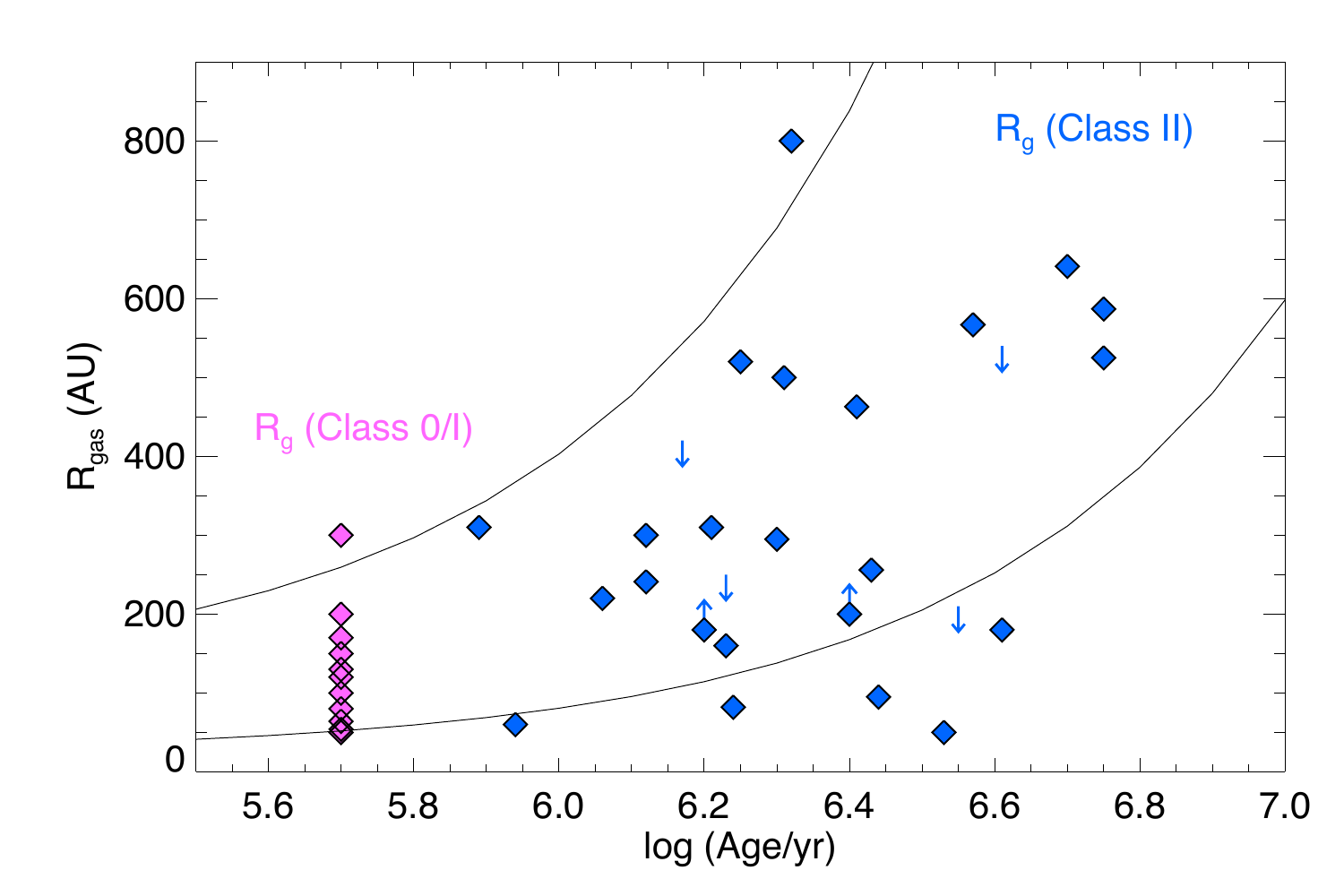}
\caption[]
{\small Observed gas disk radii of 
Class I disks (pink diamonds) and 
Taurus Class II disks (blue diamonds and arrows). 
The solid lines show the disk radii that contain 
90\% of the disk mass in simple models of viscously 
spreading disks (see text for details).
Most of the Class II gas disk sizes fall between these two 
lines.
}
\end{figure}

Figure 3 compares the gas disk sizes of Class II sources 
from Figure 2 
(blue diamonds and arrows) with those of Class I sources 
(pink diamonds). 
The latter are placed at an nominal age 
younger than 1 Myr. 
While almost all of the Class I gas disk radii are smaller than 200\,AU, 
2/3 of the Taurus Class II disk radii are larger than 200\,AU. 
If Class I disks represent the initial conditions for 
the evolution of Class II disks, 
these two properties suggest that gas disks grow in size 
in the T Tauri phase.

The Class II disk sizes are roughly consistent with the 
sizes expected for sources that start out at the sizes of 
Class I disks and spread with time as they accrete. 
To illustrate this, we can consider the evolution of a 
T Tauri ``$\alpha$-disk''.  
The observed evolution in stellar accretion rates with time 
can be explained as a consequence of 
disk evolution through a viscous process in which 
viscosity is parameterized as $\nu = \alpha c_s H$, where $c_s$ is the 
sound speed, $H$ the disk scale height, and $\alpha \simeq 0.01$ 
is the viscosity parameter 
(Hartmann et al.\ 1998; Sicilia-Aguilar et al.\ 2010). 

As described by Hartmann et al.\ (1998), 
if the viscosity varies with disk radius $R$ 
as a power law ($\nu \propto R^\gamma$), 
the viscous evolution of the disk has a similarity solution 
(Lynden-Bell \& Pringle 1974), with $\gamma =1$ 
corresponding to the usual assumption of a viscosity parameter 
$\alpha$ that is roughly constant with radius. 
In this case, the fraction of the disk mass interior to 
radius $R$ at time $t$, 
$$
{M_d(R,t)\over M_d(t)} = 1 - \exp\left(-{R\over R_1 T}\right), 
$$
where $R_1$ is the radius that initially contains 
$\sim 0.6$ of the total disk mass,  
$T$ is the non-dimensional time 
$T =  t/t_s + 1.$ 
In this expression 
$t_s$ is the viscous scaling time 
$t_s  = R_1^2/3\nu_1, $ 
where $\nu_1$ is the viscosity at $R_1,$ so that 
$$t_s \sim 0.08\,{\rm Myr}\,
\left(\alpha\over 10^{-2} \right)^{-1}
\left(R_1\over 10\,{\rm AU}\right)
\left(\Mstar\over 0.5\right)^{1/2}
\left(T_d\over 10\,{\rm K}\right)^{-1},
$$
where $T_d$ is the disk temperature at 100\,AU. 

The solid lines in Figure 3 show, as a function of time, 
the disk radii that contain 90\% of the disk mass,  
assuming 
a typical T Tauri stellar mass $\Mstar=0.5\Msun$ and 
a typical disk temperature of 10\,K at 100\,AU. 
The curves assume two different initial disk sizes and 
viscosities: 
$R_1=50$\,AU and $\alpha = 0.01$ (upper line) and 
$R_1=10$\,AU and $\alpha=0.002$ (lower line). 
The range in $R_1$ is chosen to span the sizes of 
Class I disks. 
Most of the Class II disk sizes fall between these two 
lines.

To include the Lupus sources in the comparison of 
Class I and Class II disk sizes, we also show in 
Figure 4 the size distributions of Class II 
gas disks in Taurus (blue) and Lupus (gray) 
and Class I gas disks (pink), shown 
differentially and as a cumulative 
fraction (lower right).
The cumulative size distribution of the combined 
(Taurus and Lupus) Class II gas disk samples is also 
shown (cyan; lower right). 
Like the Taurus sources, the Lupus sample is restricted 
to the stellar mass range 0.3--1.3\,$\Msun$ using the 
masses from 
Alcal\'a et al.\ (2014, 2017; 
see also online tables in Ansdell et al.\ 2018); 
sources with a known binary companion within 2\arcsec\ 
are also excluded (Sz~68 and Sz~123A; Ghez et al.\ 1997). 
For the Taurus sources with upper limits on disk mass 
(4 sources), the disk size is taken as 50\% of the upper 
limit value. The results are not sensitive to the 
exact value. 
While only 23\% of the Class I gas disks are larger than 200\,AU, 
a much larger fraction of the Taurus (68\%) and Lupus (50\%) 
Class II gas disks are larger than 200\,AU. 

From the two-sample K-S test, we find that the probability is 
$< 1$\% that the Taurus and Class I gas disk sizes 
are drawn from the same distribution. 
Similarly, the probability that the 
combined (Taurus and Lupus) Class II gas disk sizes
are drawn from the same distribution as the 
Class I gas disk sizes is also $< 1$\%. 
Because the Class I sample is small, the results are sensitive to 
which sources are included in the comparison. 
If we exclude L1551 NE (a binary with large 300\,AU gas disk)
from the Class I sample, 
the probability that the Class II and Class I sizes are drawn 
from the same distribution drops to 
$0.2$\%.

\begin{figure}[tbh!]
\figurenum{4}
\includegraphics[width=6.in]{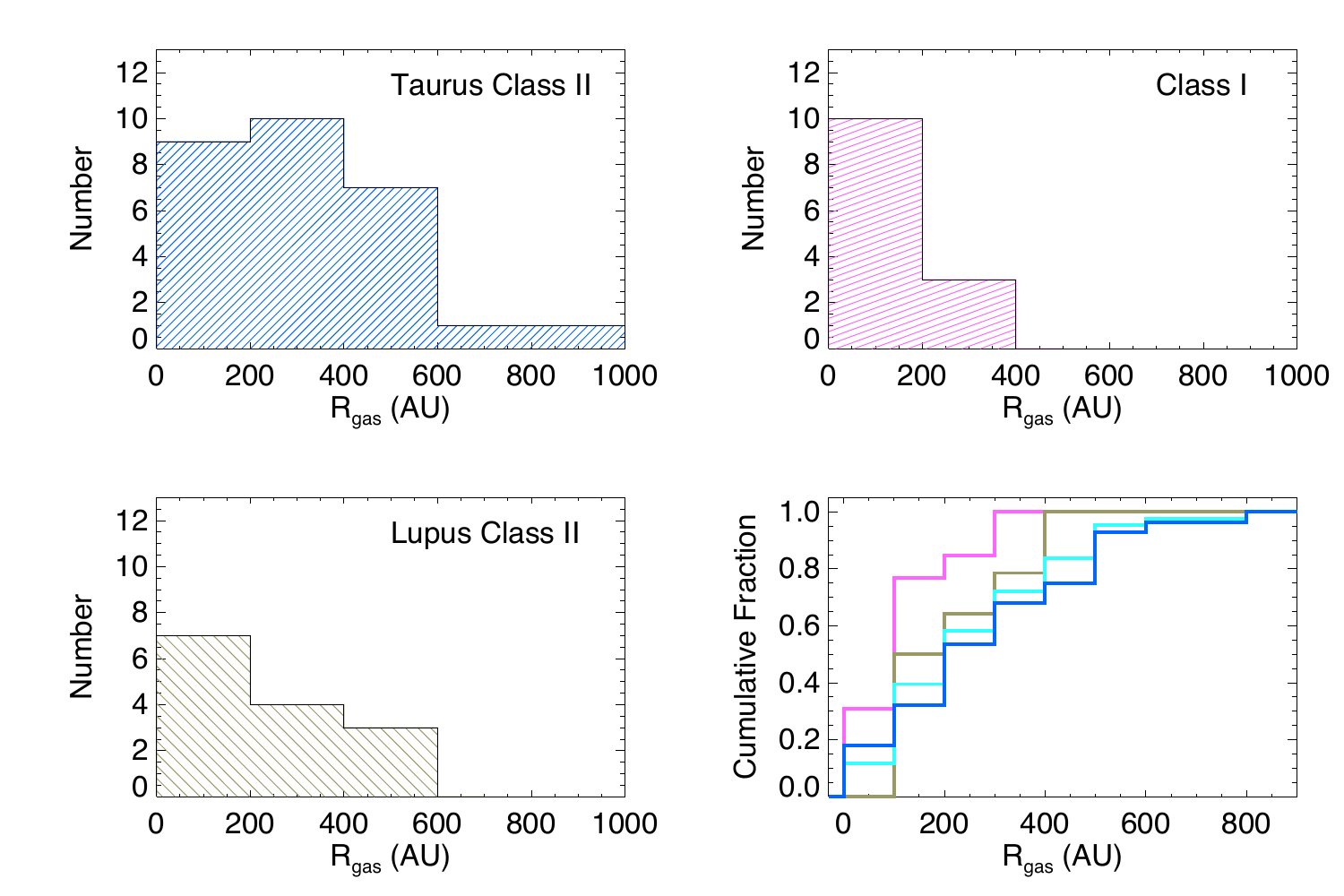}
\caption[]
{\small Gas disk sizes of Class II sources in Taurus (blue) and 
Lupus (gray) and Class I sources (pink) shown differentially 
and as a cumulative fraction (lower right).
In the lower right panel, the cumulative distribution of the 
combined (Taurus and Lupus) Class II gas disks is also shown (cyan). 
Class I gas disks are smaller than Class II gas disks. 
}
\end{figure}

\section{Discussion} 

As shown in the previous section, most Class II disks 
in Taurus 
are larger in radius than the Class I disks that have been 
studied to date, consistent with expectations for 
disk spreading, 
i.e., angular momentum 
transport within the disk in the Class II phase. 
Disks might spread as a consequence of angular momentum 
transport through gravitational instability 
(e.g., review by Kratter \& Lodato 2016) 
or viscous transport (e.g., review by Hartmann et al.\ 2016). 
Processes other than such ``in-disk'' angular momentum transport 
tend to reduce the sizes of gas disks. 
Photoevaporation by stellar FUV irradiation acts to truncate 
the disk at large radii and cause it to shrink with time, 
even in the presence of viscous spreading (Gorti et al.\ 2015). 
If magnetothermal winds (Bai \& Stone 2013; Bai et al.\ 2016) 
remove angular momentum efficiently 
from disks, disks could accrete without needing to spread with time. 
The formation of giant planets at large radii can also truncate
disks. 

Our result complements observational and theoretical studies 
of angular momentum transport in disks. 
While the MRI (Balbus \& Hawley 1991; Gammie 1996) 
has long been the favored mechanism for disk accretion 
in the T Tauri phase, recent theoretical studies that explore the 
impact of non-ideal MHD effects have seriously questioned whether 
the MRI can operate in T Tauri disks, especially at radii 
relevant to planet formation $\sim 1$--10\,AU. 
Winds launched by a combination of magnetic and thermal effects 
have been proposed as an alternative transport mechanism 
(Bai \& Stone 2013; Kunz \& Lesur 2013; Simon et al.\ 2013a, 2013b, 2015; 
Lesur et al.\ 2014; Gressel et al.\ 2015; Bai et al.\ 2016; 
Bai \& Stone 2017; 
see Turner et al.\ 2014 for a review). 

Neither mechanism (MRI or magnetothermal winds) 
has a verified observational signature thus far 
at these distances. 
Disk turbulence possibly driven by the MRI has been detected 
both within 1\,AU and beyond 40\,AU. 
At disk radii within 0.3 AU, high resolution spectroscopy of 
CO overtone emission has uncovered evidence for non-thermal 
velocities comparable to the sound speed in the disk 
atmospheres of a few young stars  
(e.g., Carr et al.\ 2004; Najita et al.\ 1996; 2009; Doppmann et al.\ 2008), 
consistent with the non-thermal motions expected 
for MRI-driven turbulence. 
High resolution ALMA observations of outer disks appear to 
favor low levels of non-thermal broadening, at only
$\sim 5$--10\% of the sound speed 
(HD163296 and TW Hya--Flaherty et al.\ 2015, 2018; 
de Gregorio-Monsalvo et al.\ 2013). 
However, turbulence at a larger fraction of the sound speed 
($\sim 20$\%) has recently been detected in the outer disk 
of DM Tau (K.\ Flaherty 2018, private communication). 
No signature of turbulence has yet been reported at the disk radii 
where non-ideal MHD effects are expected to strongly suppress the MRI 
(1--10\,AU). 

The existence and character of magnetothermal winds are also 
uncertain. Theoretical studies predict that winds capable of driving 
disk accretion at the observed stellar accretion rates will be 
massive, with mass loss rates comparable to disk accretion rates. 
It has been suggested that 
the low velocity component of the OI 6300A  line emission from 
T Tauri stars provides evidence for magnetothermal winds 
(Simon et al.\ 2016). However, the decomposition of 
a complex OI 6300A profile into multiple components potentially 
introduces uncertainty in the interpretation. 
More detailed studies of this and other diagnostics, 
combined with quantitative theoretical predictions of 
observable wind signatures can potentially verify the 
existence and angular momentum transport properties of 
magnetothermal winds.

In the meantime, 
the larger sizes of most Taurus Class II disks compared to Class I 
disks strongly suggest that angular momentum redistribution 
within the disk, by some mechanism, plays a large enough role 
in disk evolution that a large fraction of disks spread 
significantly from the Class I to Class II phases.
The data do not comment on whether the mechanism responsible 
is the MRI or other processes. 
Disk winds may also remove angular momentum but not enough to 
prevent the spreading of these disks. 

These results complement earlier commentary on the 
evolution of (primarily dust) disk sizes that 
found tentative, sometimes conflicting, results. 
Although Andrews et al.\ (2007) seemed to find no trend of 
dust disk size with age among Class II objects 
(their Fig.~15), 
subsequent studies found tentative evidence that dust disk sizes do 
increase with age (Isella et al.\ 2009, their Figure 10; 
Guilloteau et al.\ 2011, their Figure 13).  
More recently, Tazzari et al.\ (2017) found
that in star-forming regions separated by 1--2 Myr in age, 
the sizes of Class II disks, as measured from submillimeter 
{\it continuum} emission, are slightly larger for 
disks in Lupus (1--3\,Myr) 
than those in the slightly younger Taurus and Ophiuchus (1--2\,Myr) 
populations. They tentatively attributed the size difference 
to viscous evolution.  
Thus, the evidence for increasing {\it dust} disk size with age 
during the Class II phase is modest to uncertain, 
consistent with the picture from Figure 2. 

Of the 22 Lupus Class II disks in the recent study by 
Ansdell et al.\ (2018), 
which  range in size from $\sim 100$\,AU to $\sim 500$\,AU, 
approximately half have gas disk radii $>200$\,AU, 
a smaller fraction than in the Taurus sample studied 
here but still quite large. 
Although the authors did not compare the sizes of 
Class I gas disks with their Class II gas disk sizes, 
it seems clear that 
if the Class I disks from Table 1 are typical of the 
evolutionary precursors of the Lupus Class II disks, 
the large gas disks ($> 200$\,AU) among the 
Lupus population also suggest that viscous spreading 
occurs in the T Tauri phase (Figure 4). 

For the purpose of this study,  
the Lupus disks are less ideal than the Taurus disks 
for two reasons. 
Firstly, the binarity of sources in Lupus is not as 
completely characterized as that of Taurus sources. As a result, 
some Lupus disks may possess unknown stellar companions that have 
dynamically truncated their gaseous disks.
Secondly, Lupus is an older star forming region ($\sim 3$\,Myr) 
than Taurus (1--2\,Myr). 
As a result, photoevaporation has had more time to 
evaporate away outer disks (Gorti et al.\ 2015). 
Giant planets, which have had more time to form in older systems,  
can also truncate disks dynamically. 
Despite these possible effects, the Lupus Class II disks still 
appear larger than the Class I disks.

More extensive measurements of gas disk sizes are needed
to understand the timing and extent of ``in-disk'' angular 
momentum transport. 
Among the Class II disks studied here, there is a likely bias 
toward larger disks which are brighter and easier to 
study and resolve. Measurements of disk radii for a larger 
number of disks would illuminate the 
full range and frequency of gas disk sizes as a function 
of age. At the present time, 
the existence of a large number of disks that are 
larger than Class I gas sizes (Fig.\ 3) strongly suggests 
that at least some disks spread in the Class II phase. 

One of the limitations of this study is the small number of 
Class I sources with reported gas disk radii. 
The presence of an infalling envelope also makes it
challenging to measure Class I gas disk sizes; 
future work may find a way around this difficulty. 
If future studies of a larger population of 
Class I sources find disks systematically 
much larger than those studied to date, our conclusion 
will need to be revised.
Further measurements of Class I disks can also 
reveal when disk spreading occurs.
One might argue that the larger sizes of 
Class II disks are an outcome of viscous spreading in the 
Class I phase rather than the Class II phase.  
If true, surveys of a larger number of Class I disks 
should encounter Class I disks with larger sizes. 

Interestingly there is, 
in addition to the majority of large disks ($> 200$\,AU), 
a population of very small gas disks over a range of 
ages ($< 100$\,AU). The small disks shown in Figure 3 
come primarily from the 
sample studied by Simon et al.\ (2017). 
These authors suggested that the small gas disk sizes they 
measured were due in part to to the cooler effective 
temperatures of the stars in their sample. 
However, we did not find a strong trend between stellar 
luminosity and gas disk size in the sample studied here. 
Such small disks are not anticipated at ages of several Myr 
if all disks spread with an effective viscosity of $\alpha >0.001$. 
These systems may be disks that are trucated from the outside 
by photoevaporation, disk winds, or planetary companions. 
Further observations of these systems, to search for winds 
or companions, can test these ideas.

\acknowledgements
We thank the Kavli Institute for Theoretical Physics for their 
stimulating research environment and research support, 
which led us to pursue the ideas explored here.  
We also thank Megan Ansdell, Scott Kenyon, and the referee, Michal Simon, 
for helpful advice and careful readings of the manuscript. 
J.\ N.\ acknowledges the stimulating research environment 
supported by NASA Agreement No.\ NNX15AD94G to the 
``Earths in Other Solar Systems'' program. 
This research was supported in part by the National Science Foundation
under Grant No.\ NSF PHY-1748958.

\newpage

\noindent {\bf References}

\noindent Alcal\'a, J.\ M., Natta, A., Manara, C.\ F., et al.\ 2014, 
AA, 561, 2

\noindent Alcal\'a, J.\ M., Manara, C.\ F., Natta, A.\ et al.\ 2017, 
AA, 600, 20

\noindent Andrews, S.\ M., Wilner, D.\ J., Hughes, A.\ M., et al.\ 
2012, ApJ, 744, 162

\noindent Andrews, S.\ M., Rosenfeld, K.\ A., Kraus, A.\ L., \& 
Wilner, D.\ J.\ 2013, ApJ, 771, 129

\noindent Ansdell, M., Williams, J.\ P., Trapman, L., et al.\ 2018, 
ApJ, 859, 21

\noindent Antoniucci S., Garc\'ia L\'opez R., Nisini B., et al.\ 2014, 
AA, 572,62

\noindent Artymowicz, P.\ \& Lubow, S.\ H.\ 1994, ApJ, 421, 651

\noindent Aso, Y., Ohashi, N., Saigo, K., et al.\ 2015, ApJ, 812, 27

\noindent Bai, X.\ \& Stone, J.\ M.\ 2013, ApJ, 769, 76

\noindent Bai, X., Ye, J., Goodman, J., \& Yuan, F.\ 2016, ApJ, 818, 152

\noindent Bai, X., \& Stone, J. M. 2017, ApJ, 843, 36

\noindent Balbus, S.\ A., \& Hawley, J.\ F.\ 1991, ApJ, 376, 214

\noindent Birnstiel, T., \& Andrews, S. M. 2014, ApJ, 780, 153

\noindent Brauer, F., Dullemond, C.\ P., Henning, Th.\ 2008, 
AA, 480, 859

\noindent Brinch, C., \& J\/orgensen, J.\ K. 2013, AA, 559, 82

\noindent Calvet N.\ 2004, In ``Stars as Suns: Activity, Evolution
and Planets'', Proc. Symp. Int. Astron. Union, 219th, Sydney, July
21-25, 2003, ed.\ A.\ K.\ Dupree, A.\ O.\ Benz, 
(San Francisco: ASP), pp. 599-609. 

\noindent Carr, J. S., Tokunaga, A. T., \& Najita, J. 2004, ApJ, 603, 213

\noindent Chapillon, E., Guilloteau, S., Dutrey, A., \& Pi\'etu, V.\ 2008, A\&A, 488, 565

\noindent Chou, T.-L., Takakuwa, S., Yen, H.-W., Ohashi, N., \& Ho, P. T. P.\ 2014, ApJ, 796, 70

\noindent Cleeves, L. I., \"Oberg, K. I., Wilner, D. J., et al.\ 
2016, ApJ, 832, 110

\noindent Coffey, D., Dougados, C., Cabrit, S., Pety, J., \& Bacciotti, F.\ 2015, ApJ, 804, 2

\noindent de Gregorio-Monsalvo, I., M\'enard, F., Dent, W., 
et al.\ 2013, AA, 557, A133

\noindent Doppmann, G. W., Najita, J. R., \& Carr, J. S. 2008, 
ApJ, 685, 298

\noindent Dutrey, A., Guilloteau, S., Prato, L., et al.\ 1998, A\&A, 338, L63

\noindent Fang, M., van Boekel, R., Wang, W., et al.\ 2009, 
AA, 504, 461

\noindent Flaherty, K. M., Hughes, A. M., Rosenfeld, K. A., 
et al.\ 2015, ApJ, 813, 99

\noindent Flaherty, K.\ M., Hughes, A.\ M., Teague, R., et al.\ 2018, ApJ, 856, 117

\noindent Gammie, C.\ F.\ 1996, ApJ, 457, 355

\noindent Ghez, A.\ M., McCarthy, D.\ W., Patience, J.\ L., 
\& Beck, T.\ L.\ 1997, ApJ, 481, 378

\noindent Gorti, U., Hollenbach, D., \& Dullemond, C.\ P.\ 
2015, ApJ, 804, 29

\noindent Gressel, O., Turner, N.\ J., Nelson, R.\ P.\ \& McNally, C.\ P.\ 2015, 
ApJ, 801, 84

\noindent Guilloteau, S., Dutrey, A., \& Simon, M.\ 1999, A\&A, 348, 570

\noindent Guilloteau, S., Di Folco, E., Dutrey, A., et al.\ 2013, AA, 549, 92

\noindent Guilloteau, S., Simon, M., Pi\'etu, V., Di Folco, E., Dutrey, A., 
Prato, L., \& Chapillon, E.\ 2014, AA, 567, 117

\noindent Guilloteau, S., Reboussin, L., Dutrey, A., et al.\ 2016, AA, 592, 124

\noindent Guilloteau, S., Dutrey, A., Pi\'etu, V., \& Boelher, Y.\ 
2011, AA, 529, 105

\noindent Harris, R.\ J., Andrews, S.\ M., Wilner, D.\ J., \& 
Kraus, A.\ L.\ 2012, ApJ, 751, 115

\noindent Harsono, D., Visser, R., Bruderer, S., van Dishoeck, E.\ F., \& 
Kristensen, L.\ E.\ 2013, AA, 555, 45

\noindent Harsono, D., J\"orgensen, J. K., van Dishoeck, E. F., et al.\ 2014, A\&A, 562, 77

\noindent Hartmann, L., Calvet, N., Gullbring, E., \& D'Alessio, P.\ 1998, 
ApJ, 495, 385

\noindent Hartmann, L., Herczeg, G., \& Calvet, N.\ 2016, 
ARAA, 54, 135

\noindent Herczeg G.\ J., Hillenbrand, L.\ A.\ 2008, 
ApJ, 681, 594

\noindent Huang, J., Andrews, S.\ M., Cleeves, L.\ I., et al.\ 2018, ApJ, 852, 122

\noindent Isella, A., Testi, L., Natta, A., et al.\ 2007, 
AA, 469, 213

\noindent Isella, A., Carpenter, J.\ M., \& Sargent, A.\ I.\  
2009, ApJ, 701, 260

\noindent Jensen, E. L. N., \& Akeson, R.\ 2014, Nature, 511, 567

\noindent Kessler-Silacci 2004, Ph.D. Thesis

\noindent Kim, K.\ H., Watson, D.\ M., Manoj, P., et al.\ 2016, ApJS, 226, 8

\noindent Kratter, K.\ \& Lodato, G.\ 2016, ARAA, 54, 271

\noindent Kunz, M.\ W.\ \& Lesur, G.\ 2013, MNRAS, 434, 2295

\noindent Lee, C.-F., Hirano, N., Zhang, Q., et al.\ 2014, ApJ, 786, 114

\noindent Lesur, G., Kunz, M. W., \& Fromang, S. 2014, AA, 566, 56

\noindent Lin, D.\ N.\ C., Bodenheimer, P., \& Richardson, D.\ C.\ 

\noindent Liu, Y., Henning, Th., Carrasco-Gonz\'alez, C., et al.\ 2017, AA, 607, 74

\noindent Lynden-Bell, D., \& Pringle, J.\ E.\ 1974, 
MNRAS, 168, 603

\noindent Manara, C.\ F., Robberto, M., Da Rio, N., et al. 2012, 
ApJ, 755,154

\noindent Manara, C.\ F., Testi, L., Natta, A., Alcala, J.\ M.\ 2015, 
AA, 579, 66

\noindent Murillo, N. M., Lai, S.-P., Bruderer, S., Harsono, D., \& van Dishoeck, E. F.\ 2013, A\&A, 560, A103

\noindent Muzerolle, J., Hillenbrand, L., Calvet, N., et al.\ 2003, 
ApJ, 592, 266

\noindent Najita, J.\ R., Andrews, S.\ M., Muzerolle, J.\ 2015, 
MNRAS, 450, 3559

\noindent Najita, J.\ R., Strom, S.\ E., Muzerolle, J.\ 2007, 
MNRAS, 378, 369

\noindent Najita, J., Carr, J. S., Glassgold, A. E., Shu, F. H., \& 
Tokunaga, A. T. 1996, ApJ, 456, 292

\noindent Najita, J. R., Doppmann, G. W., Carr, J. S., Graham, J. R., 
\& Eisner, J. A. 2009, ApJ, 691, 738

\noindent Natta, A., Testi, L., \& Randich, S.\ 2006, AA, 452, 245
1996, Nature, 380, 606

\noindent Ohashi, N., Saigo, K., Aso, Y., et al.\ 2014, ApJ, 796, 131

\noindent Pani\'c, O., Hogerheijde, M.\ R., Wilner, D., \& Qi, C.\ 
2009, AA, 501, 269

\noindent Perez, L.\ M., Carpenter, J.\ M., Andrews, S.\ M., et al.\ 2016, Science, 353, 1519

\noindent Pi\'etu, V., Guilloteau, S., \& Dutrey, A.\ 2005, A\&A, 443, 945

\noindent Pi\'etu, V., Guilloteau, S., Di Folco, E., Dutrey, A., \& 
Boehler, Y. 2014, AA, 564, 95

\noindent Schaefer, G. H., Dutrey, A., Guilloteau, S., Simon, M., \& White, R. J.\ 2009, ApJ, 701, 698

\noindent Sicilia-Aguilar, A., Henning, T., \& Hartmann, L.\ H.\ 
2010, ApJ, 710, 597

\noindent Simon, J. B., Bai, X.-N., Stone, J. M., Armitage, P. J., \& 
Beckwith, K. 2013a, ApJ, 764, 66

\noindent Simon, J. B., Bai, X.-N., Armitage, P. J., Stone, J. M., \& 
Beckwith, K. 2013b, ApJ, 775, 73

\noindent Simon, J. B., Lesur, G, Kunz, M. W., \& Armitage, P. J.\ 2015, 
MNRAS, 454, 1117

\noindent Simon, M., Dutrey, A., \& Guilloteau, S.\ 2000, ApJ, 545, 1034

\noindent Simon, M.\ N., Pascucci, I., Edwards, S., et al.\ 2016, ApJ, 831, 169

\noindent Siess, L., Dufour, E., \& Forestini, M.\ 2000, 
AA, 358, 593

\noindent Simon, M., Guilloteau, S., Di Folco, E., et al.\ 2017, ApJ, 844, 159

\noindent Takakuwa, S., Saito, M., Lim, J., et al.\ 2012, ApJ, 754, 52

\noindent Takeuchi, T., \& Lin, D.\ N.\ C.\ 2002, 581, 1344

\noindent Takeuchi, T., \& Lin, D.\ N.\ C.\ 2005, 623, 482

\noindent Tazzari, M., Testi, L., Natta, A., et al. 2017, A\&A, 606, A88

\noindent Tomida, K., Okuzumi, S., \& Machida, M.\ N.\ et al.\ 2016, 
ApJ, 801, 117

\noindent Tripathi, A., Andrews, S.\ M., Birnstiel, T., \& Wilner, D.\ J.\ 
2017, ApJ, 845, 44

\noindent Turner, N.\ J., Fromang, S., Gammie, C., et al.\ 2014, in Protostars and Planets VI, ed. H.\ Beuther et al.\ (Tucson, AZ: Univ.\ of Arizona Press), 411

\noindent Venuti, L., Bouvier, J., Flaccomio, E., et al.\ 2014, 
AA, 570, 82

\noindent Yen, H.-W., Koch, P.\ M., Takakuwa, S., Krasnopolsky, R., 
Ohashi, N., \& Aso, Y.\ 2017, ApJ, 834 178

\noindent Zhu, Z., Hartmann, L., Gammie, C.\ F., 
Book, L.\ G., Simon, J.\ B., \& Engelhard, E.\  2010, 
ApJ, 713, 1134

\clearpage

\begin{deluxetable}{lccrlrrrl}
\tabletypesize{\footnotesize}
\tablecaption{\label{t:std} Properties of Class II Sources}
\tablehead{
Name	 & $\Mstar$ & Age & $R_d$  & Ref & $R_{\rm CN}$ & $R_g $ &e($R_g$)& Ref, Tracer \\ 
       &($M_\odot$) &(Myr)& (AU)   &     & (AU)         & (AU)   & (AU)   &      }
\startdata
      AA Tau &  0.8 & 2.0 &   400  & A07 &     300 &$\approx$ 500 &     &  K04, CO        \\
      AB Aur &  2.3 &20.9 &   ...  & ... & $<$ 250 &          890 & 10  &  P05, $^{13}$CO \\
      BP Tau &  0.8 & 1.7 &   ...  & ... & $<$ 250 &          ... &     &                 \\
     CIDA 11 &  0.4 & 4.1 &   ...  & ... & $<$ 540 &          ... &     &                 \\
      CI Tau &  0.8 & 1.8 &   233  & T17 &     ... &          520 & 13  &  P14, CN        \\
      CW Tau &  1.6 & 2.2 &    39  & P14 & $<$ 170 &          210 &  7  &  P14, $^{13}$CO \\
      CY Tau &  0.4 & 2.0 &    66  & T17 &     ... &          295 & 11  &  G14, CN        \\
      DE Tau &  0.4 & 0.9 &    19  & P14 & $<$ 250 &           60 & 20  &  P14, $^{13}$CO \\
      DL Tau &  0.8 & 2.6 &   175  & A07 &     560 &          463 &  6  &  G14, CO, CN    \\
      DM Tau &  0.5 & 5.0 &   459  & T17 &     610 &          641 & 19  &  G14, CN        \\
      DN Tau &  0.6 & 1.3 &   112  & T17 &     490 &          241 &  7  &  G14, CN        \\
      DO Tau &  0.5 & 0.8 &    44  & T17 &     310 &          ... &     &                 \\
      DQ Tau &  1.2 & 3.5 &   ...  & ... & $<$ 210 &          ... &     &                 \\
      DR Tau &  1.2 & 1.5 &   100  & A07 & $<$ 420 &          ... &     &                 \\
      DS Tau &  1.0 & 4.1 &    70  & P14 & $<$ 310 &          180 & 24  &  P14, $^{13}$CO \\
      FT Tau &  0.7 & 1.6 &   ...  & ... &     310 &          ... &     &                 \\
      GG Tau &  1.3 & 2.1 &   ...  & ... &     490 &          800 &     &  G99, $^{13}$CO \\ 
      GM Aur &  1.3 & 5.6 &   210  & T17 &     ... &          525 & 20  &  D98, CO        \\
      GO Tau &  0.6 & 5.6 &   336  & T17 &     ... &          587 & 55  &  G14, CN        \\
   Haro 6-13 &  0.6 & 1.6 &   441  & T17 & $<$ 280 &      $>$ 180 &     &  S09, CO        \\
      HK Tau &  0.5 & 2.5 &   ...  & ... &     320 &      $>$ 200 &     &  J14, CO        \\ 
     HV TauC &  0.9 & 2.7 &   139  & H12 &     310 &          256 & 51  &  G14, CN        \\ 
      IQ Tau &  0.5 & 1.1 &   131  & T17 &     560 &          220 & 15  &  G14, CN        \\
      LkCa15 &  1.0 & 3.7 &   177  & T17 &     ... &          567 & 39  &  G14, CN        \\
     MWC 758 &  1.7 & 10. &    90  & T17 & $<$ 450 &          270 & 15  &  C08, CO        \\
      RY Tau &  2.8 & 1.0 &   150  & A07 &     310 &$\approx$ 400 &     &  C15, CO        \\
      SU Aur &  2.5 & 2.9 &    13  & T17 & $<$ 140 &      $>$ 150 &     &  P14, $^{13}$CO \\
     UZ TauE &  0.7 & 1.3 &   226  & T17 &     310 &          300 & 20  &  S00, CO        \\ 
      CX Tau &  0.4 & 1.7 &    12  & P14 &     ... &          160 & 20  &  S17, CO        \\
      FM Tau &  0.6 & 3.4 &    15  & P14 &     ... &           50 &  2  &  S17, CO        \\
      IP Tau &  0.6 & 2.8 &    39  & T17 &     ... &           95 & 20  &  S17, CO        \\
      HO Tau &  0.6 &13.2 &   ...  & ... &     ... &           62 &  5  &  S17, CO        \\ 
    V710 Tau &  0.5 & 1.7 &   ...  & ... &     ... &           82 &  6  &  S17, CO        \\ 
\enddata
\vspace{-0.5cm}
\tablecomments{A07 (Andrews et al.\ 2007); 
C08 (Chapillon et al.\ 2008); 
C15 (Coffey et al.\ 2015); 
D98 (Dutrey et al.\ 1998);
G99 (Guilloteau et al.\ 1999);
G14 (Guilloteau et al.\ 2014);
H12 (Harris et al.\ 2012);
J14 (Jensen \& Akeson 2014);
K04 (Kessler-Silacci 2004);
P05 (Pi\'etu et al.\ 2005);
P14 (Pi\'etu et al.\ 2014);
S09 (Schaefer et al.\ 2009);
S00 (Simon et al.\ 2000);
S17 (Simon et al.\ 2017);
T17 (Tripathi et al.\ (2017)
}
\end{deluxetable}

\end{document}